\documentclass[pdflatex,sn-nature]{sn-jnl}

\usepackage{graphicx}
\usepackage{multirow}
\usepackage{amsmath,amssymb,amsfonts}
\usepackage{amsthm}
\usepackage{mathrsfs}
\usepackage[title]{appendix}
\usepackage{xcolor}
\usepackage{textcomp}
\usepackage{manyfoot}
\usepackage{booktabs}
\usepackage{algorithm}
\usepackage{algorithmicx}
\usepackage{algpseudocode}
\usepackage{listings}

\theoremstyle{thmstyleone}

\theoremstyle{thmstyletwo}

\theoremstyle{thmstylethree}

\begin{document}

\title[Eco-efficiency as a Catalyst for Citizen Co-production]{Eco-efficiency as a Catalyst for Citizen Co-production: Evidence from Chinese Cities}

\author*[1]{\fnm{Ruiyu} \sur{Zhang}}\email{ruiyu.zhang@polyu.edu.hk}
\author[1]{\fnm{Lin} \sur{Nie}}\email{lin-apss.nie@polyu.edu.hk}
\author[2]{\fnm{Ce} \sur{Zhao}}\email{Alcor\_zhao@outlook.com}
\author[1]{\fnm{Xin} \sur{Zhao}}\email{xinnn.zhao@connect.polyu.hk}

\affil*[1]{\orgdiv{Department of Applied Social Sciences}, \orgname{The Hong Kong Polytechnic University}, \orgaddress{\city{Hong Kong}, \country{China}}}
\affil[2]{\orgdiv{School of Sciences}, \orgname{Harbin Institute of Technology Shenzhen}, \orgaddress{\city{Shenzhen}, \state{Guangdong}, \country{China}}}

\abstract{We examine whether higher eco-efficiency encourages local governments to co-produce environmental solutions with citizens. Using Chinese provincial data and advanced textual analysis, we find that high eco-efficiency strongly predicts more collaborative responses to environmental complaints. Causal inference suggests that crossing a threshold of eco-efficiency increases co-production probabilities by about 24 percentage points, indicating eco-efficiency’s potential as a catalyst for participatory environmental governance.}


\maketitle

\section{Introduction}\label{sec1}

As governments all over the world struggle to balance economic growth with environmental stewardship, eco-efficiency has emerged as a promising metric to align these objectives \cite{Ge2021,Stern2000}. It involves creating more economic value while minimizing resource inputs and ecological harm, offering both a guiding principle and a practical tool for policymakers  \cite{Seppala2005,Caiado2017}. Although eco-efficiency’s ability to reduce emissions and foster cost savings has been well documented \cite{Grossman1995,Wu2020}, much less is known about its potential to shape governance behaviors—particularly whether it incentivizes local officials to collaborate with citizens on shared environmental challenges \cite{Huppes2005a,Huppes2009,Koskela2012,Zhao2023}. 

While eco-efficiency's impacts on environmental and economic outcomes are increasingly documented, its potential to shape governance—particularly local authorities' willingness to engage citizens in collective problem solving—remains understudied.

One avenue for such collaboration is co-production, whereby governments and citizens jointly design or improve public services\cite{Ostrom1996, Osborne2016,Osborne2021}.In the context of the environment, co-production can leverage community insights, strengthen accountability, and accelerate policy implementation \cite{Miller2020}. Yet few studies have systematically examined whether higher eco-efficiency promotes this participatory model of governance.

In this paper, we investigate whether local eco-efficiency “spills over” to increase government willingness to engage citizens in solving environmental problems. We focus on 31 Chinese provinces, where rapid industrialization often jeopardizes sustainability mandates \cite{Wu2020}. Using Data Envelopment Analysis (DEA) techniques to measure eco-efficiency\cite{banker1984some}, we analyze over 4,000 environmental complaints and official responses. We then deploy an advanced machine-learning method (BERT, XGBoost, SHAP)\cite{Devlin2019,chen2016xgboost,lundberg2020local} and a Causal Effect Variational Autoencoder (CEVAE) to estimate the impact of eco-efficiency on co-production\cite{lundberg2020local}. Our findings shed new light on the broader governance implications of eco-efficiency and can inform global efforts to prioritize ecological gains alongside inclusive policymaking\cite{Ma2023}.

\section{Measurement and Classification}\label{sec2}

We measured province-level eco-efficiency with DEA results, designating emissions and environmental indicators as inputs and GDP as the output \cite{Caiado2017}. Provinces were then classified as high or low eco-efficiency. Next, we collected over 4,000 citizen–government environmental-related complaints and official replies across China. Each official response was manually coded as “co-production oriented” (inviting collaboration) or “one way” (regulatory or perfunctory) \cite{Osborne2016,Miller2020}. We performed spectral clustering on BERT embeddings to group complaint texts into eight thematic categories (see Figure 1).

\begin{figure}[h]
\centering
\includegraphics[width=0.9\textwidth]{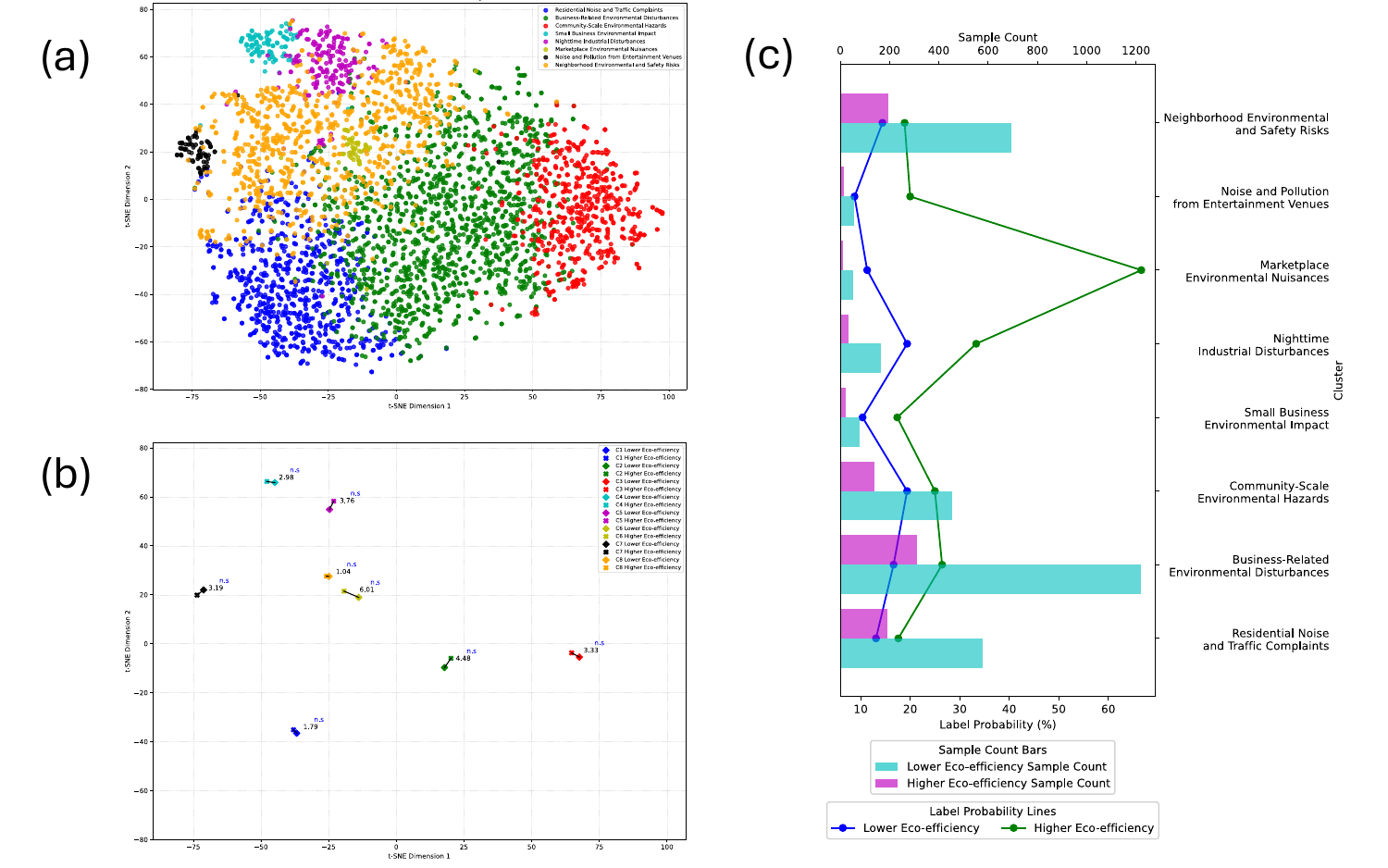}
\caption{\textbf{Clustering and Co-production Analysis of Citizens' Messages.} (a) Each point represents a single citizen message, color coded as one of eight environmental complaint clusters. These clusters were derived via spectral clustering from BERT CLS embedding methods. (b) Centroid positions for each cluster, split by high vs. low eco-efficiency provinces. The minimal distance between centroids suggests that citizens in different eco-efficiency contexts report similar types of concerns. (c) The chart shows the probability of receiving a co-production-oriented government response for each complaint cluster.}\label{fig1}
\end{figure}

\section{Predicting Co-production}\label{sec3}

To determine which factors most strongly predict a co-production-oriented response, we trained an XGBoost classifier using provincial-level attributes (e.g. eco-efficiency, budget expenditures, emissions) and complaint-level features (e.g. topic cluster, presence of high levels of public attention). We also employed an advanced large-language modeling program (Qwen, frequently used to analyze Chinese content) to assess the intensity of the messages’ sentiment to control for the impact of text sentiment on government responses. The model’s validation accuracy of 0.8922 and a test accuracy of 0.8946 indicate a reliable predictive capability.

A SHAP interpretability analysis revealed that eco-efficiency ranked among the top predictors of co-production (see figure 2a). Provinces with higher eco-efficiency tended to provide more collaborative responses, all else being equal. Other influential factors included total fiscal budget, public attention, and the perceived severity of the complaint. Certain expenditures—like agriculture and forestry spending—were slightly more negatively associated with co-production in some provinces, perhaps reflecting a preference for centralized environmental interventions rather than citizen-driven initiatives\cite{Huppes2005a,Koskela2012}.

\begin{figure}[h]
\centering
\includegraphics[height=0.5\textheight]{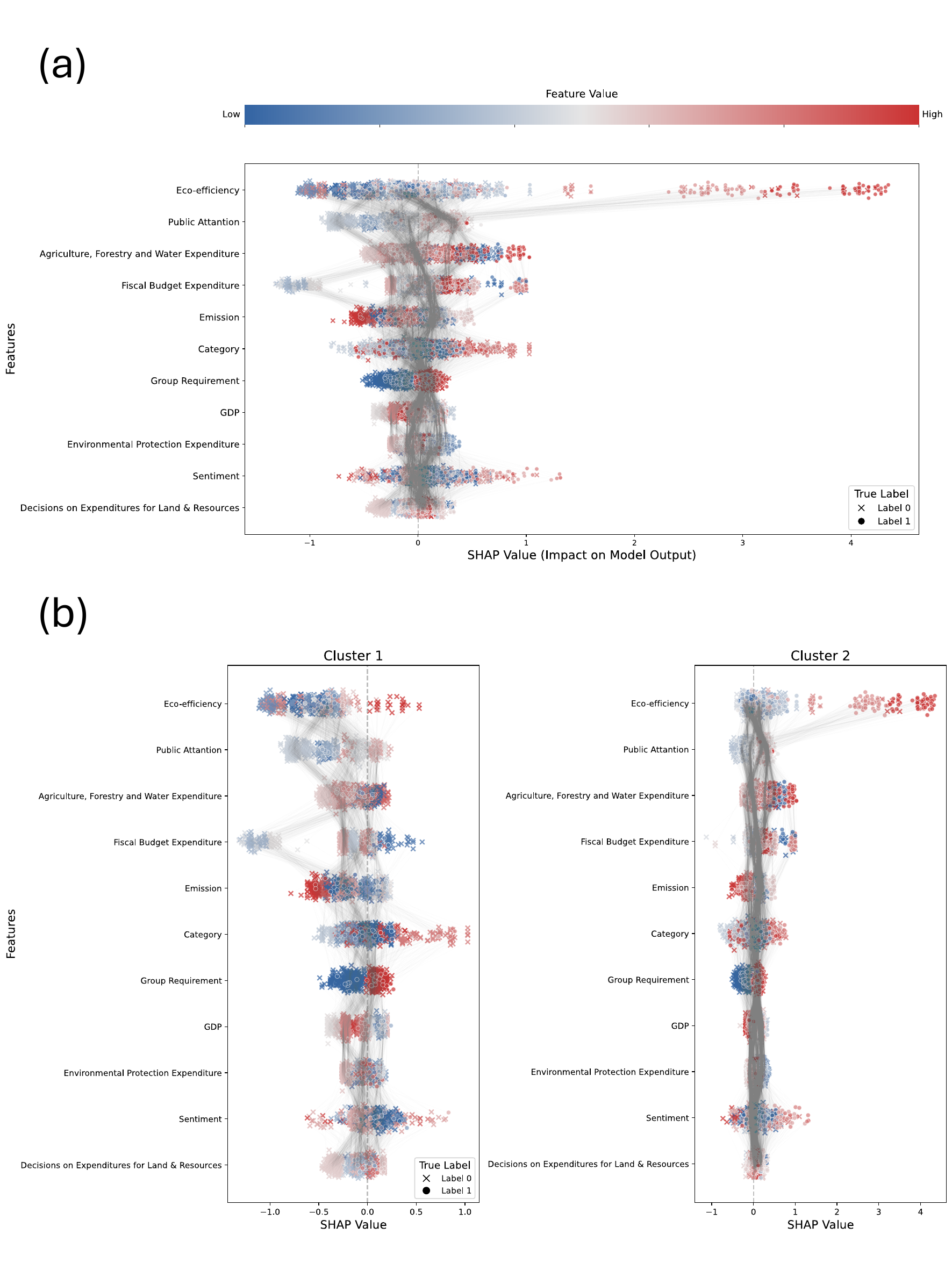}
\caption{\textbf{SHAP Analysis of Co-Production Prediction.} (a) SHAP Plot of Co-Production Prediction. Each point corresponds to a single prediction; blue (red) points indicate lower (higher) feature values. The x-axis depicts the SHAP value's impact on predicting a co-production-oriented response (label 1) versus a one-way response (label 0). The figure indicates that eco-efficiency is a particularly influential predictor; higher values push the model toward co-production classification. (b) Cluster-Specific SHAP Analysis. These SHAP summary plots compare two clusters identified by k-means. The relative importance and direction (positive or negative SHAP value) of each feature differ between clusters, highlighting how eco-efficiency and resource allocation patterns drive distinct co-production behaviors.}\label{fig2}
\end{figure}

 subsequent k-means clustering of province-level co-production probabilities yielded two distinct governance archetypes (figure 2b). Cluster 1 (less eco-efficient provinces, left panel) responded more reactively to citizen concerns, limiting opportunities for collaborative problem solving. Cluster 2 (more eco-efficient localities, right panel) featured systematically higher co-production engagement—even after controlling for population size, average income, and baseline levels of public activism\cite{Osborne2021}.

\section{Causal Inference via CEVAE}\label{sec4}

While the above findings suggest a robust correlation between eco-efficiency and co-production, endogeneity concerns remain. For instance, politically progressive provinces might invest simultaneously in green innovation and citizen engagement, generating spurious associations. To address these concerns, we employed a CEVAE, designating high eco-efficiency as the “treatment.” CEVAE models the joint distribution of treatment, covariates, and outcomes (co-production likelihood) to estimate counterfactual scenarios for each province.

The CEVAE analysis yielded an average treatment effect (ATE) of approximately 0.24, with a 95\% confidence interval of [0.2207, 0.2902]. Put simply, crossing the high-eco-efficiency threshold increases a province’s predicted probability of co-production-oriented responses by roughly 24 percentage points. These results reinforce the notion that eco-efficiency can “spill over” beyond environmental gains to influence how governments interact with citizens.

\section{Discussion and Conclusion}\label{sec5}

Our findings highlight that eco-efficiency iss a potentially powerful driver of co-production willingness. In line with earlier theoretical arguments \cite{Ostrom1996,Osborne2016,Osborne2021}, we observe that higher eco-efficiency frees up fiscal and administrative capacity, reduces environmental stressors, and bolsters public trust—which in turn encourages collaborative governance. Eco-efficient provinces may thus be better equipped to engage citizens in mitigating ongoing challenges such as air pollution or neighborhood health risks \cite{Ma2023}.

These results have broad implications beyond China. Developed regions seeking to meet ambitious net-zero targets might harness eco-efficiency gains to deepen community engagement and foster local buy-in for green transitions. Emerging economies, which likely experience acute constraints on public budgets and regulatory enforcement, could benefit from adopting eco-efficiency to bolster institutional momentum and capacity. Over time, improved environmental metrics may reduce conflictual relationships between officials and residents, creating space for joint problem solving on issues like waste management, water quality, or urban planning \cite{Wu2020}.

Yet, the mere presence of eco-efficiency does not automatically guarantee participatory decision-making. Less eco-efficient provinces often face tight resource constraints; local officials might struggle to implement even rudimentary pollution controls, let alone convene citizen consultations. They may need to engage in targeted investment in resource optimization, alongside institutional reforms to enable co-creation, to break cycles of limited civic trust and reactive governance \cite{Seppala2005}.

This study demonstrates that eco-efficiency can act as more than an environmental or economic benchmark: it may catalyze more participatory, citizen-centered governance. By applying DEA to classify localities by their levels of eco-efficiency, BERT to assess government responses, XGBoost-SHAP to identify key predictors, and CEVAE to tease out causal inference, we conclude that high eco-efficiency significantly increases co-production-oriented governance. Such “spill-over” effects resonate globally as municipalities strive to reconcile ecological imperatives with fiscal constraints and public demands for inclusive decision-making.

On a practical level, policymakers might pair resource optimization measures (e.g., emissions reduction, green infrastructure) with explicit co-production pathways, incentivizing local agencies to partner with residents. In this way, eco-efficiency improvements can yield not only smaller environmental footprints but also stronger civic trust and more durable governance solutions. Further research can deepen our understanding of this synergy by tracking eco-efficiency over time, exploring diverse governance systems, and examining how emerging concepts like circular economy or net-zero transitions influence citizen engagement. Ultimately, recognizing eco-efficiency’s potential to foster participatory governance offers a hopeful perspective on how localities can tackle complex environmental challenges while empowering their communities.

\section*{Methods}\label{sec6}
Our study integrates economic, environmental, and textual data to establish causal links between eco-efficiency and citizen-government co-production. We analyzed 27 Chinese provinces/municipalities from 2018, leveraging follow methodological approaches:
\newline

\textbf{Data collection:} We compiled provincial-level data on economic indicators, environmental metrics, and governance characteristics from official Chinese statistical yearbooks. We collected 4,221 citizen-government text exchanges concerning environmental issues from standardized provincial government online platforms, covering all mainland provinces. 
\newline

\textbf{Eco-efficiency calculation:} We employed input-oriented Data Envelopment Analysis (DEA) with variable returns to scale following Charnes et al. \cite{banker1984some} using the R package \texttt{Benchmarking}. The formulation of our input-oriented VRS DEA model is as follows:

\begin{align}
\min_{\theta, \lambda} \quad & \theta \\
\text{subject to} \quad & \sum_{j=1}^{n} \lambda_j x_{ij} \leq \theta x_{io}, \quad i = 1,2,\ldots,m \\
& \sum_{j=1}^{n} \lambda_j y_{rj} \geq y_{ro}, \quad r = 1,2,\ldots,s \\
& \sum_{j=1}^{n} \lambda_j = 1 \\
& \lambda_j \geq 0, \quad j = 1,2,\ldots,n
\end{align}

where $\theta$ is the efficiency score; $x_{ij}$ and $y_{rj}$ are the environmental inputs and GDP output, respectively; $\lambda_j$ are the weights; and the constraint $\sum_{j=1}^{n} \lambda_j = 1$ imposes variable returns to scale. The model incorporated environmental consumption as input variables (emissions and environmental indicators) and GDP as the output variable, generating eco-efficiency scores that measure how efficiently provinces convert environmental resources into economic value. We then categorized provinces into high and low eco-efficiency groups using the median score as the threshold, creating balanced treatment and control groups.
\newline

\textbf{Text analysis:} We employed a two-pronged approach to analyze citizen-government interactions. First, we manually coded all government responses as either "co-production-oriented" (explicitly inviting citizen collaboration) or "one-way" (regulatory or perfunctory). For citizen complaints, we utilized a pre-trained Chinese BERT model (\texttt{bert-base-chinese} from the \texttt{transformers} library; \cite{Devlin2019}) to transform each message into vector representations (CLS embeddings). We then applied spectral clustering (\texttt{sklearn.cluster.SpectralClustering}) to these embeddings to identify thematic patterns in environmental concerns. The spectral clustering algorithm can be formalized as follows:

\begin{align}
\mathbf{L} &= \mathbf{D} - \mathbf{W} \\
\mathbf{L}_{\text{norm}} &= \mathbf{D}^{-1/2}\mathbf{L}\mathbf{D}^{-1/2}
\end{align}

where $\mathbf{W}$ is the similarity matrix of the embeddings, $\mathbf{D}$ is the diagonal degree matrix with $D_{ii} = \sum_j W_{ij}$, and $\mathbf{L}_{\text{norm}}$ is the normalized Laplacian matrix. We then computed the $k$ eigenvectors corresponding to the $k$ smallest eigenvalues of $\mathbf{L}_{\text{norm}}$ and applied $k$-means clustering to these eigenvectors.

The optimal number of clusters ($k=8$) was determined using the Elbow method, with distinct categories emerging for issues. Permutation tests confirmed the stability and significance of these clusters ($p$ $< 0.001$), following:

\begin{align}
p = \frac{\sum_{i=1}^{N} \mathbf{1}(s_i \geq s_{\text{obs}})}{N}
\end{align}

where $s_{\text{obs}}$ is the observed clustering score, $s_i$ is the score for the $i$-th permutation, $N$ is the number of permutations, and $\mathbf{1}(\cdot)$ is the indicator function.
\newline

\textbf{Predictive modeling:} We trained an XGBoost gradient boosting classifier (\texttt{xgboost.XGBClassifier}; \cite{chen2016xgboost}) with 5-fold cross-validation to identify factors predicting co-production responses. The model incorporated 27 features including eco-efficiency scores, fiscal expenditures across sectors, environmental indicators, and complaint-specific characteristics. We employed SHAP (SHapley Additive exPlanations) values (\texttt{shap}; \cite{lundberg2020local}) to quantify each feature's contribution to predictions. 
\newline

\textbf{Causal inference:} To address potential endogeneity and estimate causal effects, we implemented a Causal Effect Variational Autoencoder (CEVAE) (\texttt{pyro.contrib.cevae.CEVAE}; \cite{Louizos2017}) using PyTorch. This deep learning approach utilizes latent-variable modeling to account for unobserved confounders. We designated high eco-efficiency as the treatment condition, incorporating 7 observed covariates and estimating a 20-dimensional latent confounder. The model architecture consisted of 3 hidden layers with 200 units each. We validated the approach through sensitivity analyses and robustness checks, comparing results with meta-learner frameworks (S-learner, T-learner, X-learner, and R-learner), which confirmed the robustness of our findings across different causal estimation strategies.

\section*{Declarations}

\bmhead{Funding}
This research was not supported by any grant.

\bmhead{Competing interests}
The authors declare no competing interests.

\bmhead{Ethics approval}
Ethical approval was obtained from the ethics committee at all relevant universities.

\bmhead{Data availability}
The datasets generated and analyzed during the current study are available from the corresponding author upon reasonable request.

\bibliography{references}

\end{document}